\documentclass[amsmath, amssymb, aps, reprint, superscriptaddress, nofootinbib]{revtex4-2}

\usepackage{graphicx}
\usepackage{dcolumn}

\usepackage{bm}
\usepackage{physics}
\usepackage[svgnames]{xcolor}
\usepackage[colorlinks=true, linkcolor=blue, citecolor=blue, urlcolor=blue]{hyperref}
\usepackage{amsthm}
\usepackage{enumitem}

\begin{document}

\title{Deep learning of thermodynamic laws from microscopic dynamics}

\author{Hiroto Kuroyanagi}
\affiliation{Department of Physics, Shizuoka University, Shizuoka 422-8529, Japan}

\author{Tatsuro Yuge}
\affiliation{Department of Physics, Shizuoka University, Shizuoka 422-8529, Japan}

\begin{abstract}
  We numerically show that a deep neural network (DNN) can learn macroscopic thermodynamic laws purely from microscopic data. Using molecular dynamics simulations, we generate the data of snapshot images of gas particles undergoing adiabatic processes. We train a DNN to determine the temporal order of input image pairs. We observe that the trained network induces an order relation between states consistent with adiabatic accessibility, satisfying the axioms of thermodynamics. Furthermore, the internal representation learned by the DNN act as an entropy. These results suggest that machine learning can discover emergent physical laws that are valid at scales far larger than those of the underlying constituents---opening a pathway to data-driven discovery of macroscopic physics.
\end{abstract}

\maketitle

\section{Introduction}

In recent years, significant advances in machine learning (ML), particularly deep neural networks (DNNs) \cite{Goodfellow_etal2016,Bishop2024}, have led to many studies applying these methods to physics.
One such area of studies is the discovery of physical laws from data using ML.
Symbolic regressions have successfully rediscovered symbolic expressions of many physical laws from data \cite{UdrescuTegmark2020,Udrescu_etal2020,Keren_etal2023,Cornelio_etal2023,MakkeChawla2024}.
ML-based modelings of dynamical equations have succeeded in predicting the time evolution of systems from time-series data \cite{Greydanus_etal2019,Choudhary_etal2020,Cranmer_etal2020,Han_etal2021}.
Also, several works have proposed ML algorithms that find hidden symmetries and conservation laws \cite{Wetzel_etal2020,mototake2021,LiuTegmark2021,LiuTegmark2022}.
These approaches typically aim to discover laws that are valid at the same scale as the data---microscopic laws from microscopic data or macroscopic laws from macroscopic data.
It remains elusive whether ML can discover physical laws emergent at a macroscopic level from microscopic data
though there are several studies of ML-based finding of appropriate latent variables whose dimension is significantly reduced compared with that of the data \cite{Champion_etal2019,Iten_etal2020,mototake2021,UdrescuTegmark2021,Chen_etal2022}.
A typical example of emergent macroscopic laws is thermodynamics \cite{Callen1985,LiebYngvason1999,LiebYngvason1998,Thess2011}.
Given only the data of dynamics of microscopic elements, it is significantly difficult to find the laws of thermodynamics.

Another area of studies is ML-based analysis of thermodynamics and statistical mechanics.
Earlier work \cite{TorlaiMelko2016} used Boltzmann machines to learn the probability distribution of microscopic states from the data of spin configurations generated by Monte Carlo simulation in equilibrium situations.
In this line of research, supervised and unsupervised MLs are used to detect phase transitions from the data of microscopic states in several systems \cite{Wang2016,OhtsukiOhtsuki2016,Carrasquilla2017,Broecker_etal2017,TanakaTomiya2017,Wetzel2017,WetzelScherzer2017,Chng_etal2018,Iso_etal2018,Canabarro_etal2019,Dong_etal2019,Zhang_etal2019,Aoki_etal2019,Kashiwa_etal2019,FunaiGiataganas2020,MiyajimaMochizuki2023,Bayo_etal2025,MochizukiMiyajima2025}.
In many of these studies, thermodynamic quantities such as order parameters are explicitly extracted in latent variables or implicitly encoded in weights of neural networks.
However, these studies do not pursue the ML-based discovery of thermodynamic laws from microscopic data.

In this paper, we demonstrate that we can construct a DNN in which thermodynamics laws are encoded by training with the data of microscopic elements.
As microscopic data, we use images (extremely short videos) of particles generated by molecular dynamics (MD) simulation of gas particles.
The DNN is supervisedly trained with the microscopic data to classify which microscopic image corresponds to the later-time state.
We numerically show that the classification of the trained DNN induces an order relation that satisfies the axioms of thermodynamics.
Moreover, the representation of an input microscopic image extracted by the trained DNN corresponds to an entropy of the state.

A related work is Ref.~\cite{Seif_etal2021}, in which a binary classification between forward and backward trajectories is used to determine the arrow of time.
The authors of Ref.~\cite{Seif_etal2021} have shown that the entropy production is extracted in the representation of the optimal neural network.
In their study the input is the whole trajectories of small systems (a single Brownian particle and few spins) in contact with a thermal bath and driven by an external agent, whereas the input of our study is an image at a certain time of a larger system of gas particles in an adiabatic process.
Also, our demonstration is not only finding the entropy in the representation of the neural network but also automatic encoding of the axioms of thermodynamics.

\section{Brief review of thermodynamics}
\label{sec:thermodynamics}

To clarify what would demonstrate that a DNN has learned thermodynamic laws, we briefly review the axiomatic framework of thermodynamics developed by Lieb and Yngvason \cite{LiebYngvason1999} (see also Refs.~\cite{LiebYngvason1998,Thess2011}).

We consider a thermodynamic system
and its state space $\Gamma$ consisting of macroscopic equilibrium states of the system, where states in $\Gamma$ may have different energies.
Composition of two states $X \in \Gamma_1$ and $Y \in \Gamma_2$ is denoted by $(X, Y)$, where $X$ and $Y$ are simply placed side by side without mixing or thermal contact.
Scaling by factor $\lambda$ of a state $X \in \Gamma$ is denoted by $\lambda X$.
When $X$ is specified by energy $U$, volume $V$, and number of particles $N$,
$\lambda X$ is an equilibrium state that is specified by $\lambda U$,  $\lambda V$, and  $\lambda N$.
If $\lambda < 1$, in particular, $\lambda X$ is obtained by taking only a fraction $\lambda$ of $X$.
The important concept in the framework is the adiabatic accessibility from $X$ to $Y$---symbolically expressed by $X \prec Y$---which means that there exists an adiabatic process that transforms $X$ to $Y$.
We write $X \sim Y$ if $X \prec Y$ and $Y \prec X$.
Also, we write $X \prec\prec Y$ if $X \prec Y$ but $Y \not\prec X$.
We say $X$ and $Y$ are comparable if either $X \prec Y$ or $Y \prec X$ holds.

The framework assumes the following basic axioms \cite{LiebYngvason1999}:
\begin{enumerate}[label=(A\arabic{*})]
  \item Reflectivity:
        $X \sim X$.
        \label{A1:reflectivity}
  \item Transitivity:
        $X \prec Y$ and $Y \prec Z$ implies $X \prec Z$.
        \label{A2:transitivity}
  \item Consistency:
        $X \prec Y$ and $X' \prec Y'$ implies $(X, X') \prec (Y, Y')$.
        \label{A3:consistency}
  \item Scaling invariance:
        If $X \prec Y$, then $\lambda X \prec \lambda Y$ for all $\lambda > 0$.
        \label{A4:scaling}
  \item Splitting and recombination:
        For $0 < \lambda < 1$, $X \sim \bigl( \lambda X, (1 - \lambda) X \bigr)$.
        \label{A5:splitting}
  \item Stability:
        For two states $X$ and $Y$,
        if $(X, \varepsilon_k Z_0) \prec (Y, \varepsilon_k Z_1)$ holds for a sequence of factors $\varepsilon_k$ tending to zero and some $Z_0, Z_1$,
        then $X \prec Y$.
        \label{A6:stability}
\end{enumerate}
In a gas system, splitting corresponds to the insertion of an adiabatic partition into the system, while recombination corresponds to the removal of the partition.
It is also assumed that any two states in the same state space are comparable.
This is called the Comparison Hypothesis and is eventually derived from the above and additional axioms in Ref.~\cite{LiebYngvason1999}.

The central result of Ref.~\cite{LiebYngvason1999} is to prove that from these axioms and the Comparison Hypothesis, one can construct an essentially unique thermodynamic entropy $S$,
which satisfies the monotonicity, $X \prec Y$ if and only if $S(X) \le S(Y)$,
and extensivity, $S(\lambda X) = \lambda S(X)$, and additivity, $S(X,Y) = S(X) + S(Y)$.
In constructing the thermodynamic entropy $S$, Ref.~\cite{LiebYngvason1999} introduces a canonical entropy $ S_\Gamma$ for a state space $\Gamma$ as
\begin{equation}
  S_\Gamma(X | X_{*}, X_{**})
  = \sup \bigl\{ \lambda \mid \bigl( (1 - \lambda) X_{*}, \lambda X_{**} \bigr) \prec X \bigr\},
  \label{canonical_entropy}
\end{equation}
where $X_{*}, X_{**} \in \Gamma$ are reference states that satisfy $X_{*} \prec\prec X_{**}$.
The two reference states serve to define the origin and scale of the canonical entropy.
Change of reference states yields an affine transformation of canonical entropy:
$S_\Gamma(X \mid X'_{*}, X'_{**}) = \alpha S_\Gamma(X \mid X_{*}, X_{**}) + \beta$,
which corresponds to a shift of the origin and a rescaling of the unit.
The canonical entropy satisfies the monotonicity when $X$ and $Y$ are in the same state space.
Although the canonical entropy $S_\Gamma$ is defined separately within each state space $\Gamma$, constructing a consistent thermodynamic entropy across different state spaces requires aligning their respective origins and scales.
In fact, it is shown in Ref.~\cite{LiebYngvason1999}
that there exists a pair of reference states (and a resulting $\alpha, \beta$) for every state space,
such that the canonical entropy can be regarded as the thermodynamic entropy for a family of state spaces, which satisfies the monotonicity, extensivity, and additivity.

Let us also introduce a useful conceptual tool: the Lieb--Yngvason machine \cite{Thess2008,Thess2011}.
It is a device that determines whether or not $X_0 \prec X_1$ holds,
based on the order relation $\prec$ satisfying \ref{A1:reflectivity}--\ref{A6:stability}.
The operation of a Lieb--Yngvason machine is expressed by the following binary-valued function $F_{\mathrm{LY}}$ of two comparable states $X_0, X_1$:
\begin{align}
  F_{\mathrm{LY}}(X_0, X_1) =
  \begin{cases}
    1 & (X_0 \prec X_1)
    \\
    0 & (\text{otherwise}).
  \end{cases}
  \label{eq:LiebYngvasonMachine}
\end{align}
In terms of the Lieb--Yngvason machine, the canonical entropy $S_\Gamma(X | X_{*}, X_{**})$ in Eq.~\eqref{canonical_entropy} is identified with $\lambda$ at which the value of
$f(\lambda) \equiv F_{\mathrm{LY}}\bigl( \bigl( (1 - \lambda) X_{*}, \lambda X_{**} \bigr) , X \bigr)$
jumps from one to zero.

From the axiomatic thermodynamics,
we specify our goal for demonstrating that a DNN can learn thermodynamic laws:
Can we make a DNN machine that behaves like a Lieb--Yngvason machine
by training it with microscopic data and without prior knowledge of thermodynamics?

\section{Preparation of data: MD simulation}
\label{sec:md}

We use MD simulation to prepare the dataset in this study.
This is because we would like to demonstrate that a DNN can learn thermodynamics from microscopic physics
and the training data should obey the microscopic laws (classical mechanics in this study).
Moreover, since we would like to train a DNN without prior knowledge of thermodynamics and statistical mechanics,
we do not use any thermostats or barostats in the simulation.
For the same reason, we do not use Monte Carlo simulation in contrast to the previous studies \cite{TorlaiMelko2016,Wang2016,OhtsukiOhtsuki2016,Carrasquilla2017,Broecker_etal2017,TanakaTomiya2017,Wetzel2017,WetzelScherzer2017,Chng_etal2018,Iso_etal2018,Canabarro_etal2019,Dong_etal2019,Zhang_etal2019,Aoki_etal2019,Kashiwa_etal2019,FunaiGiataganas2020,MiyajimaMochizuki2023,Bayo_etal2025,MochizukiMiyajima2025}
(Monte Carlo simulation samples data from canonical ensemble, so that it contains quantities specific to thermodynamics).

\begin{figure}[t]
  \includegraphics[width=\linewidth]{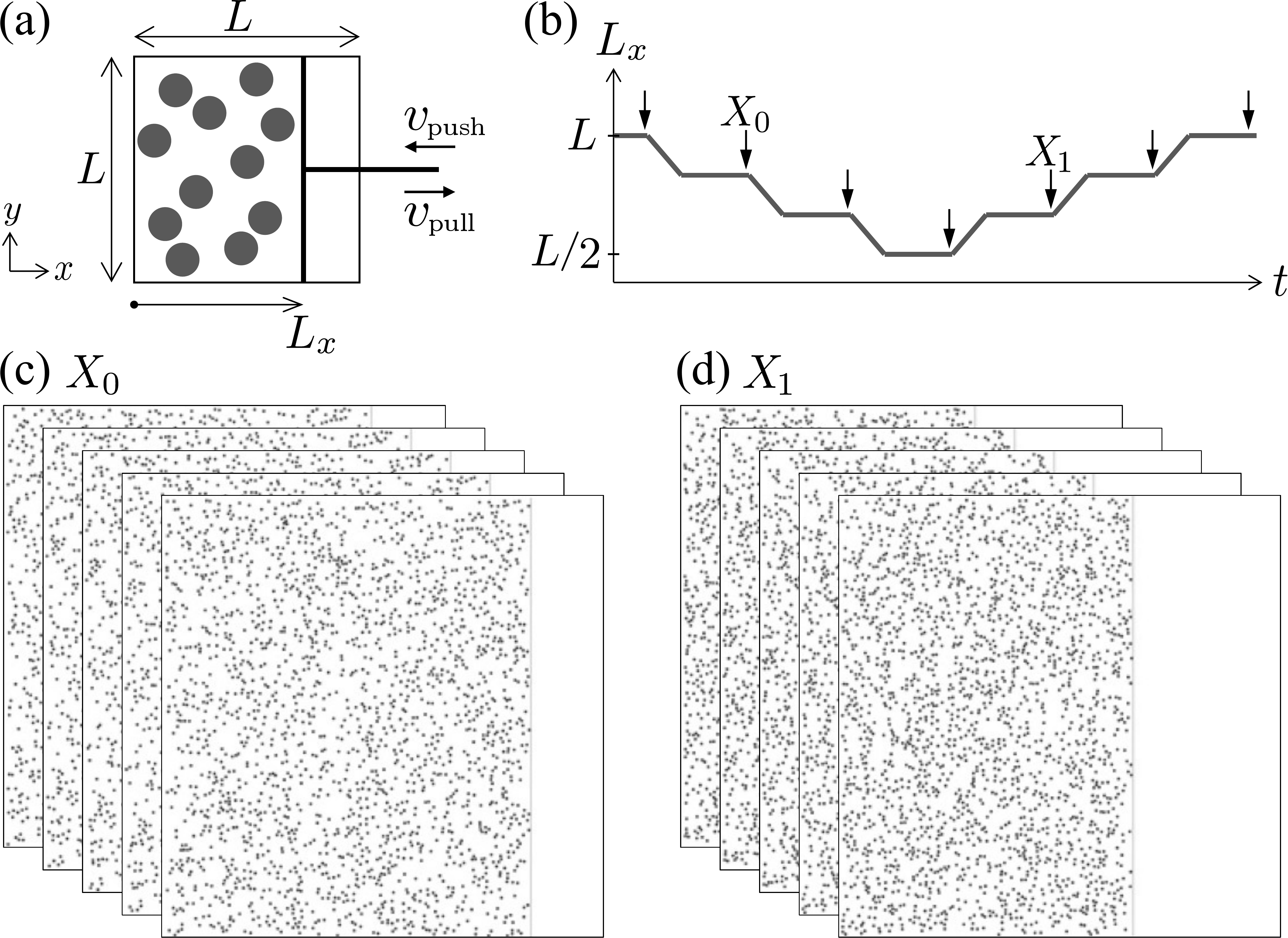}
  \caption{(a) Schematic illustration of the MD simulation model.
    (b) An example of the time profile of the piston position $L_x$ for a single cycle.
    Five snapshots (an extremely short video) are taken at each of the times indicated by the down arrows.
    Snapshots in (c) and (d) are actual examples obtained from the MD simulation for the case of $N=2000$, $v_{\mathrm{push}} = v_{\mathrm{pull}} = 0.1$, and $\Delta L_{\mathrm{push}} = \Delta L_{\mathrm{pull}} = 50$.
    The time points at which the snapshots were taken are indicated in (b) by $X_0$ for (c) and $X_1$ for (d).
  }
  \label{fig:md}
\end{figure}

We consider a two-dimensional system of $N$ particles in a thermally insulated container of size $L \times L$, as schematically depicted in Fig.~\ref{fig:md}(a).
We have a piston on the right side of the container and can control its position $L_x$ and speed $v_{\mathrm{push}/\mathrm{pull}}$, as shown in Fig.~\ref{fig:md}(b).

We assume a short-range repulsive interaction between the particles. Specifically, the interaction potential between the $i$th and $j$th particles (at the positions of $\bm{r}_i$ and $\bm{r}_j$, respectively) is
\begin{align}
  \phi(\bm{r}_i, \bm{r}_j) =
  \begin{cases}
    \frac{1}{2}C_{\mathrm{int}} \bigl( 2R - r_{i,j} \bigr)^2
      & (r_{i,j} \le 2R)
    \\
    0 & (r_{i,j} > 2R),
  \end{cases}
  \label{particle_potential}
\end{align}
where the positive constant $C_{\mathrm{int}}$ represents the interaction strength, $R$ is the radius of each particle, and
$r_{i,j} = |\bm{r}_i - \bm{r}_j|$.
We also assume that the repulsive interaction potential $\phi_{\mathrm{wall}}$ between the particles and the container, or the piston, have functional forms similar to that of Eq.~\eqref{particle_potential}.
The energy of the system in a state $X$ is given by
\begin{align}
  U(X)
  = \sum_{i=1}^N \left[ \frac{1}{2}m |\bm{v}_i|^2 + \phi_{\mathrm{wall}}(\bm{r}_i) \right]
  + \sum_{i<j} \phi(\bm{r}_i, \bm{r}_j),
  \label{energy_md}
\end{align}
where $\bm{r}_i$ and $\bm{v}_i$ are the position and velocity of the $i$th particle in the state $X$.

The initial positions of the particles are randomly arranged with the condition of no overlap.
The initial velocities of the particles are sampled from the Maxwell distribution with temperature $T_0$.
In the simulation, we express physical quantities in units of energy $k_{\mathrm{B}} T_0$, length $R$, and time $R \sqrt{m / k_{\mathrm{B}} T_0}$, where $k_{\mathrm{B}}$ is the Boltzmann constant and $m$ is the particle mass.
We set $L = 300$ and $C_{\mathrm{int}} = 10000$ in these units and $N = 500, 1000, 1500, 2000$ for individual simulation runs.
We adopt the velocity Verlet algorithm \cite{Swope_etal1982} for integrating the equations of motion in time $t$, where the time-step width is 0.001.

We control the piston as follows.
Initially, we set it at $L_x = L$.
After a while, we begin to push it at a constant speed $v_{\mathrm{push}}$.
When we move it a certain distance $\Delta L_{\mathrm{push}}$, we stop and hold it for a while.
We repeat these push and stop periods until it reaches $L_x = L/2$.
Then we reverse the direction by repeating pull and stop periods until it returns to $L_x = L$,  where the speed of the piston is $v_{\mathrm{pull}}$
(which may be different from $v_{\mathrm{push}}$)
and the distance moved in a period is $\Delta L_{\mathrm{pull}}$
(which may be different from  $\Delta L_{\mathrm{push}}$).
In a single run of the MD simulation, we repeat this cycle of $L_x = L \to L/2 \to L$ [Fig.~\ref{fig:md}(b)] three times.
We refer to an entire sequence composed of these three cycles as a ``process.''
We note that a process is thermodynamically adiabatic because the container and piston are the potential walls and no thermostats are used.
We also note that the density (packing fraction) $\eta = N \pi R^2 /(L L_x)$ is smaller than 0.14 in our simulation,
which implies that the system is in the dilute gas phase and far from the Alder-type transition point ($\eta \sim 0.7$) \cite{AlderWainwright1962,Engel_etal2013}.

\begin{figure}[t]
  \includegraphics[width=\linewidth]{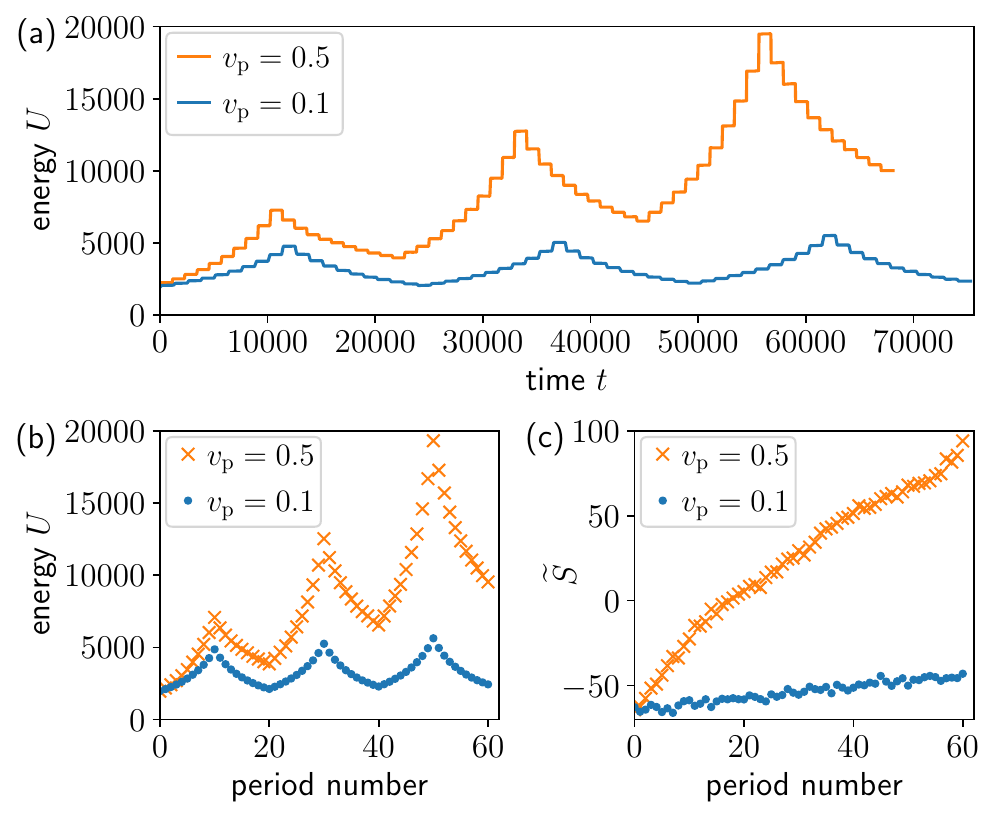}
  \caption{
    (a) Time evolution of the energy $U$ obtained from the MD simulation for cases of $v_{\mathrm{push}} = v_{\mathrm{pull}} = v_{\mathrm{p}}$.
    (b) Energy $U$ at the end of each piston-stopping period [each of the flat regions in (a)], plotted against the period number.
    (c) Neural network output $\tilde{S}$ for the state at the end of each piston-stopping period.
    Results for $v_{\mathrm{p}} = 0.5$ and $v_{\mathrm{p}} = 0.1$ are plotted.
    The other parameters are
    $N=2000$ and $\Delta L_{\mathrm{push}} = \Delta L_{\mathrm{pull}} = 15$.
  }
  \label{fig:profiles}
\end{figure}

We perform MD simulation for 448 processes.
Each of the processes has a different set of values of $N \in \{ 500, 1000, 1500, 2000 \}$, $v_{\mathrm{push/pull}} \in \{ 0.1, 0.5 \}$, and $\Delta L_{\mathrm{push/pull}} \in \{150, 75, 50, 37.5, 30, 25, 18.75,15, 10, 7.5 \}$.
In Fig.~\ref{fig:profiles}(a),
we plot the time profiles of energy $U$ for processes with faster ($v_{\mathrm{push}} = v_{\mathrm{pull}} = 0.5$) and slower ($v_{\mathrm{push}} = v_{\mathrm{pull}} = 0.1$) operations of the piston.
We observe that the energy is not monotonic in time due to the fact that the process involves both adiabatic compression (work being done on the system) and adiabatic expansion (work being done by the system).
We also observe that the energy change is larger in the faster operation than that in the slower.
Since the mean speed of a particle in the initial state is $\sqrt{2 k_{\mathrm{B}} T_0 / m} = \sqrt{2}$ (from the law of equipartition of energy),
$v_{\mathrm{push/pull}} = 0.5$ is not so slow that the faster operation is far from the quasi-static ones.

At around the end of each stop period in a process,
we record the positions of the particles five times, with a time interval of $\Delta t = 1$ between each recording.
At each recording time, we create a snapshot image of $1560 \times 1560$ pixels by drawing circles of radius $R$ at the recorded positions and compress it to a $256 \times 256$ pixel image using the Lanczos filter.
For every stop period, we thus obtain a series of five images, which composes an extremely short video and which explicitly encodes the particle positions and implicitly encodes their velocities through frame-to-frame displacements.
Hereafter, we refer to a single series of the five images simply as a ``microscopic image'' or ``microscopic state.''
In Figs.~\ref{fig:md}(c) and (d), we show examples of microscopic images for states $X_0$ and $X_1$ indicated in Fig.~\ref{fig:md}(b).

At the end of this section, we briefly remark on the connection between microscopic time-reversal symmetry and macroscopic irreversibility. Since the MD simulations in this study are conducted in closed systems, the dynamics are microscopically time reversible.
This raises the question: if two states $X$ and $Y$ are connected through a single process, does it imply that both the forward evolution from $X$ to $Y$ and the reverse evolution from $Y$ to $X$ are allowed, and that the two states are adiabatically accessible from each other (i.e., $X \sim Y$)?
A modern perspective on this issue is provided by the fluctuation theorems (see e.g., Refs.~\cite{Campisi_etal2011,Sagawa2012}):
although both time evolutions are microscopically possible, when initial states are sampled from an equilibrium ensemble, the probability of one is overwhelmingly smaller than that of the other. In macroscopic systems, this results in only one direction of the process being realized with almost certainty.
Therefore, it is expected that the data from the MD simulations include information on macroscopic irreversibility and that a DNN can learn the ordering of adiabatic accessibility from the data.

\section{Neural network and supervised learning}
\label{sec:nn}

As discussed at the end of Sec.~\ref{sec:thermodynamics},
our goal is to make a DNN machine that behaves like a Lieb--Yngvason machine
with microscopic data and without prior knowledge of thermodynamics.
To achieve this goal, it is important to appropriately design both the task and the architecture of the DNN.

First, we consider the task.
As described in Eq.~\eqref{eq:LiebYngvasonMachine}, the Lieb-Yngvason machine takes a pair of states as input and outputs a binary decision of the adiabatic accessibility between them.
To obtain such a machine through training, it is therefore natural to formulate a binary classification task that takes two states as input.
However, directly training a model to classify pairs based on adiabatic accessibility would require explicit labels for such accessibility, which are not available in the data and would contradict the aim of learning without prior thermodynamic knowledge.
Instead, we note that the MD simulation data represent time evolution under adiabatic processes.
Motivated by the macroscopic irreversibility under adiabatic processes, we hypothesize that the temporal order of states may reflect their adiabatic accessibility.
Thus, we propose to train a model to infer the temporal order between two states.
This leads to the following binary classification task:
Given a pair of microscopic images sampled from a single process in Sec.~\ref{sec:md}
[see an example of a pair ($X_0, X_1$) in Fig.~\ref{fig:md}],
which of the two input states occurred later in the process?
In this task, each sample in the dataset consists of $(X_0, X_1; y)$,
where the input variables $X_0$ and $X_1$ are two microscopic images
and the target variable $y \in \{0, 1\}$ represents the image label of the later state (i.e., $X_y$ is the later state) in the actual MD simulation.

\begin{figure}[t]
  \includegraphics[width=\linewidth]{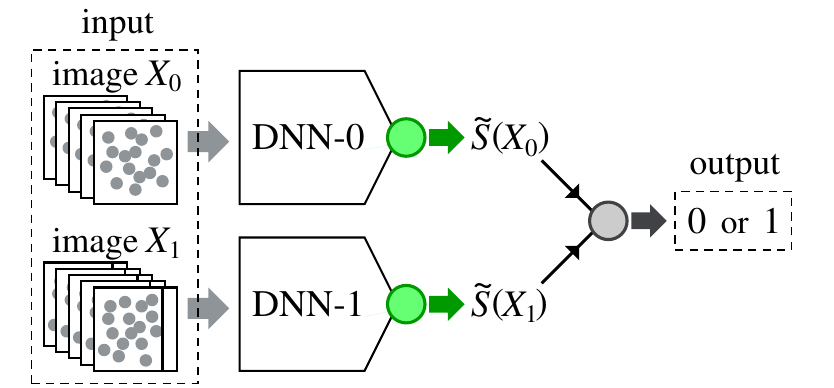}
  \caption{Schematic illustration of the DNN model.
    The input is a pair of microscopic states $X_0$ and $X_1$, each of which is an image having $256 \times 256$ pixels and five channels.
    The five channels correspond to the snapshots composing an extremely short video.
    Each of the two subnetworks DNN-0 and DNN-1 outputs a representation $\tilde{S}$ of the input image.
    The two representations are fed forward to the sigmoid unit in the output layer of the whole network to compute the probability of $X_1$ being the later state.
    The output is the label of the predicted later state.
  }
  \label{fig:dnn}
\end{figure}

Next, we consider the architecture of the DNN designed to solve the task defined above.
The input to the model consists of a pair of system states, and the output is a label predicting which of the two corresponds to the later time step.
To produce this output, it is natural to extract a set of features from each input state and compare them to infer the temporal order.
This approach motivates the use of a Siamese neural network \cite{Bromley_etal1993,Chopra_etal2005,Bertinetto_etal2016,Wetzel_etal2020,Li_etal2022},
a class of models commonly used to assess similarity between pairs of inputs.
A Siamese network comprises two identical subnetworks that share parameters; each subnetwork processes one of the inputs and extracts its features.
The outputs of the two subnetworks are then compared to determine the final prediction.
A key design choice is the dimensionality of the feature representations produced by each subnetwork.
Guided by the second law of thermodynamics---which implies a monotonic increase in entropy over time in an adiabatic process---we hypothesize that a single scalar feature per input suffices to capture the temporal ordering.
We expect that this scalar can be interpreted as a learned proxy for entropy, enabling the model to reflect the expected monotonic trend when comparing state pairs.
We also note that since each input corresponds to a sequence of images, a convolutional neural network (CNN) is a natural choice for the architecture of each subnetwork within the Siamese framework.

These considerations lead to the overall architecture of our DNN, as illustrated in Fig.~\ref{fig:dnn}.
It consists of two subnetworks (denoted by DNN-0 and DNN-1), which are CNNs sharing the parameters (weights and biases).
DNN-0 (DNN-1) takes a microscopic image $X_0$ ($X_1$) as an input and outputs a single scalar representation $\tilde{S}(X_0)$ ($\tilde{S}(X_1)$) of the image,
where the detailed structure of the subnetwork is shown in Table~\ref{tab:dnn}.
These representations are fed forward to the neuron in the output layer of the overall network.
It computes the probability $p$ that $X_1$ is the later state by $p = f_{\mathrm{sigmoid}}\bigl( \tilde{S}(X_1) - \tilde{S}(X_0) + \epsilon \bigr)$
and outputs 0 if $p < 1/2$ and 1 otherwise,
where $f_{\mathrm{sigmoid}}(x) = 1 / (1 + e^{-x})$ is the logistic sigmoid function
and the small constant $\epsilon=10^{-4}$ is added in the argument for forcing the DNN to output 1 if $\tilde{S}(X_1) = \tilde{S}(X_0)$.
We express the overall operation of this DNN architecture as $F_{\mathrm{DNN}}(X_0, X_1)$,
which is a binary-valued function of two states $X_0, X_1$.

\begin{table}[t]
  \caption{Structure of a subnetwork in the DNN model, which corresponds to DNN-0 or DNN-1 in Fig.~\ref{fig:dnn}.
    All convolutional layers have filters of size $2 \times 2$ with a stride of $2$.
  }
  \label{tab:dnn}
  \begin{ruledtabular}
    \begin{tabular}{l@{\hspace{2em}}l@{\hspace{2em}}l}
      Layer         & Hyperparameter       & Activation \\
      \colrule
      Convolutional & Output channels: 16  & ReLU       \\
      Convolutional & Output channels: 32  & ReLU       \\
      Convolutional & Output channels: 64  & ReLU       \\
      Convolutional & Output channels: 64  & ReLU       \\
      Convolutional & Output channels: 128 & ReLU       \\
      Flatten       & -                    & -          \\
      Dropout       & Rate: 0.25           & -          \\
      Dense         & Output units: 128    & ReLU       \\
      Dense         & Output units: 64     & ReLU       \\
      Dense         & Output unit: 1       & Linear     \\
    \end{tabular}
  \end{ruledtabular}
\end{table}

We have thus constructed the DNN model $F_{\mathrm{DNN}}$ that has the inputs and output similar to those of the Lieb--Yngvason machine $F_{\mathrm{LY}}$.
We note that the inputs $X_0, X_1$ of $F_{\mathrm{DNN}}$ are microscopic states, which are specified by the particle positions and velocities,
whereas those of $F_{\mathrm{LY}}$ are macroscopic states, which are specified by a set of extensive variables (energy $U$, volume $V$, and number of particles $N$ for the system of this study).

We train the DNN with the cross-entropy loss function and the Adam optimizer with a decreasing learning rate \cite{Goodfellow_etal2016,Bishop2024}.
Among the 448 processes in the MD simulation, 336 and 56 processes are used for training and validation, respectively, and the remaining 56 processes are used for test.
To the training data we apply data augmentation consisting of image rotation (by angle $\pi/2$, $\pi$ and $3\pi/2$), image flip (in the x- and y-directions), time-reverse of video, and their combinations.
We use TensorFlow \cite{tensorflow2015} to implement and train the DNN.

After training, we evaluate the performance of the DNN on the test dataset. The test loss is 0.1041, and the test accuracy reaches 0.9534. In Fig. 2(c), we show the learned representation $\tilde{S}$ for states in processes involving faster and slower piston operations. We observe that $\tilde{S}$ increases almost monotonically for both processes, whereas the energy $U$ does not [Fig. 2(b)]. Moreover, the increasing trend of $\tilde{S}$ is more pronounced in the faster operation.
These observations suggest that changes in energy alone cannot account for the high classification accuracy, while $\tilde{S}$ captures macroscopic irreversibility through its monotonic increase---implying an entropic nature of $\tilde{S}$. In the next section, we explore this point in more detail.

\section{Results}
\label{sec:results}

\subsection{Axiomatic properties of the DNN}
\label{subsec:axioms_dnn}

We now investigate whether the trained DNN machine have learned the thermodynamic laws.
We define the order relation $\prec_{\mathrm{DNN}}$ induced by the machine:
$X \prec_{\mathrm{DNN}} Y$ if $F_{\mathrm{DNN}}(X, Y) = 1$.
In the following, using the test dataset, we demonstrate that $\prec_{\mathrm{DNN}}$ satisfies the axioms \ref{A1:reflectivity}--\ref{A6:stability} within the numerical precision.
This demonstration implies that the machine behaves like a Lieb--Yngvason machine.
We note that by definition, $X \prec_{\mathrm{DNN}} Y$ if and only if $\tilde{S}(X) \le \tilde{S}(Y)$.
Also, we write $X \sim_{\mathrm{DNN}} Y$
if both $X \prec_{\mathrm{DNN}} Y$ and $Y \prec_{\mathrm{DNN}} X$ hold.

\begin{figure}[tb]
  \centering
  \includegraphics[width=.75\linewidth]{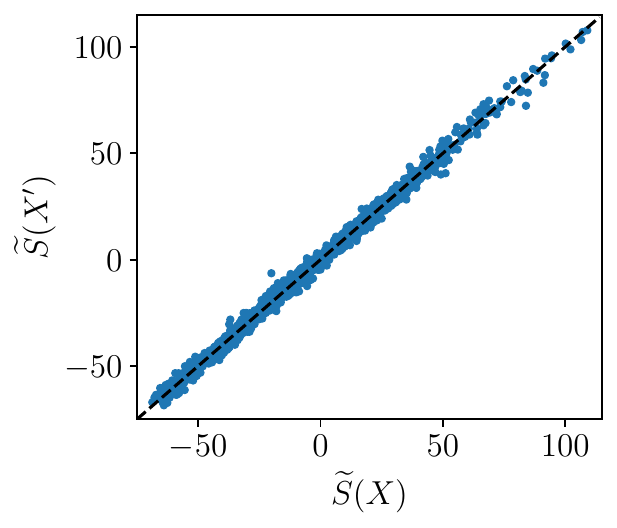}
  \caption{
    Numerical demonstration of \ref{A1:reflectivity}.
    The representation $\tilde{S}$ of a state $X'$ macroscopically equivalent to $X$ is plotted against $\tilde{S}(X)$.
    The dashed line shows $\tilde{S}(X') = \tilde{S}(X)$.
  }
  \label{fig:axiom1}
\end{figure}

First, we consider the reflectivity \ref{A1:reflectivity}.
By the definition of the DNN machine, it is obvious that $X \sim_{\mathrm{DNN}} X$ holds.
However, we note that $\sim$ and $\prec$ are the relations between macroscopic states
whereas $\sim_{\mathrm{DNN}}$ and $\prec_{\mathrm{DNN}}$ are the ones between microscopic states.
Therefore, to demonstrate \ref{A1:reflectivity},
we must examine whether $X \sim_{\mathrm{DNN}} X'$ holds if $X$ and $X'$ are macroscopically equivalent.
We also note that confirming $\tilde{S}(X) = \tilde{S}(X')$ with small error is sufficient to numerically demonstrate $X \sim_{\mathrm{DNN}} X'$.
To create a state $X'$ macroscopically equivalent to $X$,
we perform another MD simulation run of the same process as that of $X$, starting with a microscopically different but macroscopically equivalent initial state (with the same initial values of energy, volume and particle number).
As seen in Fig.~\ref{fig:axiom1}, $\tilde{S}(X) \simeq \tilde{S}(X')$ holds for most of the test data.
In fact, the coefficient of determination $R^2$ of the plot in Fig.~\ref{fig:axiom1} is $0.9951$, which is near to 1 and therefore numerically demonstrates \ref{A1:reflectivity}.

Second, we consider the transitivity \ref{A2:transitivity}.
This is obviously valid because
$X \prec_{\mathrm{DNN}} Y$ and $Y \prec_{\mathrm{DNN}} Z$
imply $\tilde{S}(X) \le \tilde{S}(Y) \le \tilde{S}(Z)$ and thus $X \prec_{\mathrm{DNN}} Z$.

\begin{figure}[tb]
  \centering
  \includegraphics[width=.96\linewidth]{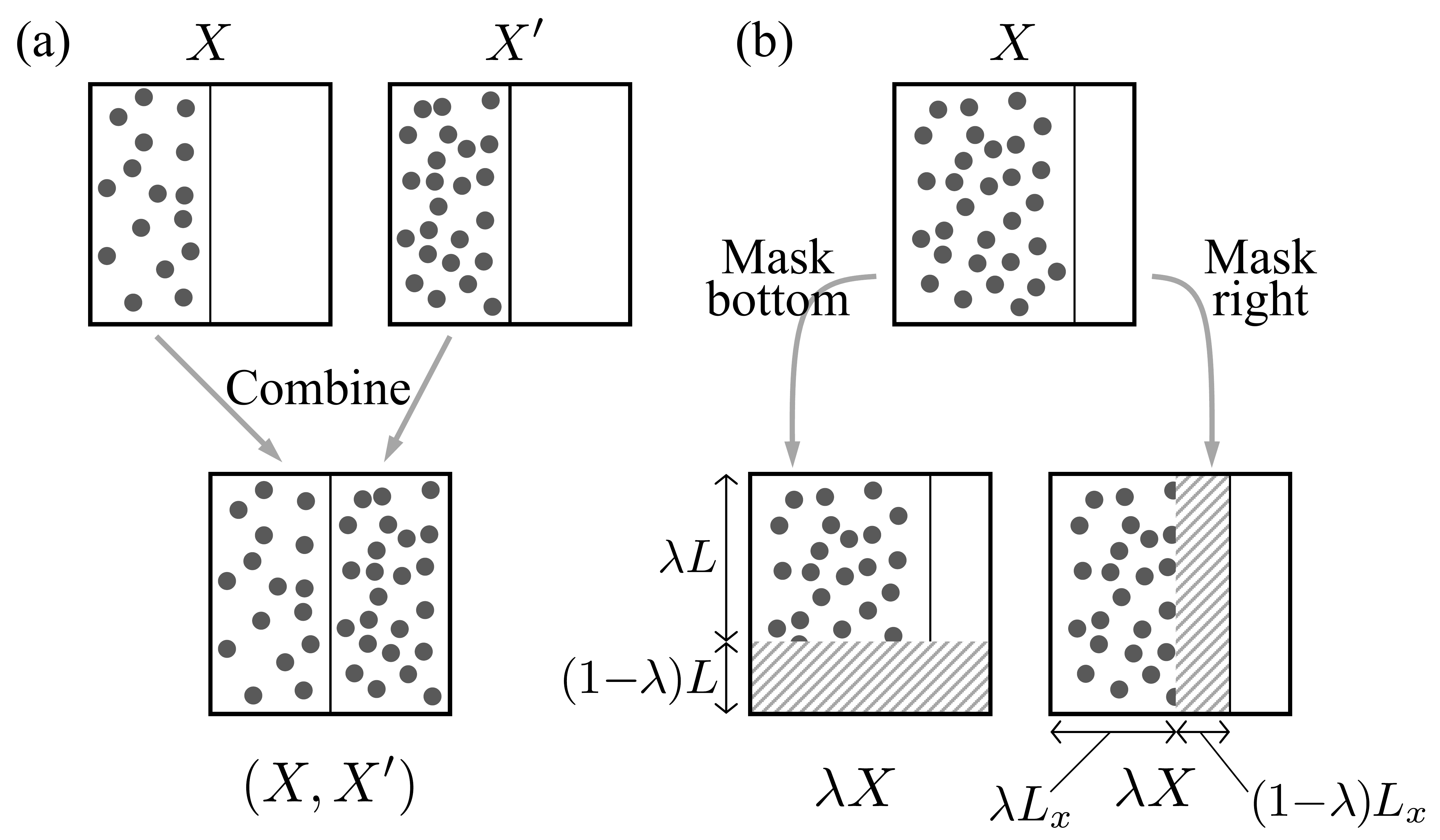}
  \caption{
    (a) Illustration of creating a compound state $(X,X')$.
    The images of two states, $X$ and $X'$, where the pistons are at $L_x=L/2$ are combined into a single image of $(X,X')$.
    (b) Illustration of creating a scaled state $\lambda X$.
    Either the bottom or right part (shaded) in the image of a state $X$ is masked.
    In the actual data, the masked region is filled with zero.
  }
  \label{fig:compound_scaling}
\end{figure}

Third, we consider the consistency \ref{A3:consistency}.
We create a compound state $(X, X')$ as follows [Fig.~\ref{fig:compound_scaling}(a)]:
We take two states $X$ and $X'$ where the pistons are located at $L_x = L/2$, cut out the half of each of the two images where the particles are confined, and combine them to create a single image of $(X, X')$.
We note that this state $(X, X')$ can be interpreted as an equilibrium state of a compound system, where $X$ and $X'$ are simply placed side by side without mixing or thermal contact.
For such compound states, we investigate the prediction agreement rate, defined by $| D_{\mathrm{A3}}^{\mathrm{agree}} | / | D_{\mathrm{A3}} |$,
where
$D_{\mathrm{A3}} = \{(X, X', Y, Y') \mid X \prec_{\mathrm{DNN}} Y, X' \prec_{\mathrm{DNN}} Y', L_x(X)=L_x(X')=L_x(Y)=L_x(Y')=L/2 \}$
and
$D_{\mathrm{A3}}^{\mathrm{agree}} = \{(X, X', Y, Y') \in D_{\mathrm{A3}} \mid (X, X') \prec_{\mathrm{DNN}} (Y, Y') \}$
in the test dataset,
and $|D|$ represents the number of data in a dataset $D$.
We find that the prediction agreement rate is 0.9995.
That is, if $X \prec_{\mathrm{DNN}} X'$ and $Y \prec_{\mathrm{DNN}} Y'$, this DNN model can predict $(X, X') \prec_{\mathrm{DNN}} (Y, Y')$ with the accuracy of 99.95\%, which demonstrates \ref{A3:consistency}.

Fourth, we consider the scaling invariance \ref{A4:scaling}.
To create a scaled state $\lambda X$, we mask the $(1-\lambda)$ fraction of either the bottom or right region of the microscopic image $X$  [Fig.~\ref{fig:compound_scaling}(b)], where we use five different values of $\lambda$ ($= 0.5, 0.6, 0.7, 0.8, 0.9$).
The scaled state thus obtained corresponds to taking a part (unmasked region) of the system. Strictly speaking, because particles can cross the boundary between the masked and unmasked regions, the unmasked region is not a perfectly closed system. However, the length of an input video (of five frames) is only $5 \times \Delta t = 5$, during which the change of particle number is negligible.
Thus, for practical purposes it can be regarded as a closed subsystem.
For such scaled states, we investigate the prediction agreement rate, $(1/5) \sum_\lambda | D_{\mathrm{A4}}^{\mathrm{agree}}(\lambda) | / | D_{\mathrm{A4}} |$ with
$D_{\mathrm{A4}} = \{(X, Y) \mid X \prec_{\mathrm{DNN}} Y \}$
and
$D_{\mathrm{A4}}^{\mathrm{agree}}(\lambda) = \{(X, Y) \in D_{\mathrm{A4}} \mid \lambda X \prec_{\mathrm{DNN}} \lambda Y \}$
in the test dataset.
We find that the prediction agreement rate is 0.9417, which demonstrates \ref{A4:scaling} with the accuracy of 94.17\%.
The prediction agreement rate of \ref{A4:scaling} is lower than that of \ref{A3:consistency}.
This is possibly because the number of particles $N(\lambda X)$ in the scaled state (created by masking) is not exactly the same as $\lambda N(X)$ and the deviation $N(\lambda X) - \lambda N(X)$ fluctuates sample by sample.

\begin{figure}[tb]
  \centering
  \includegraphics[width=\linewidth]{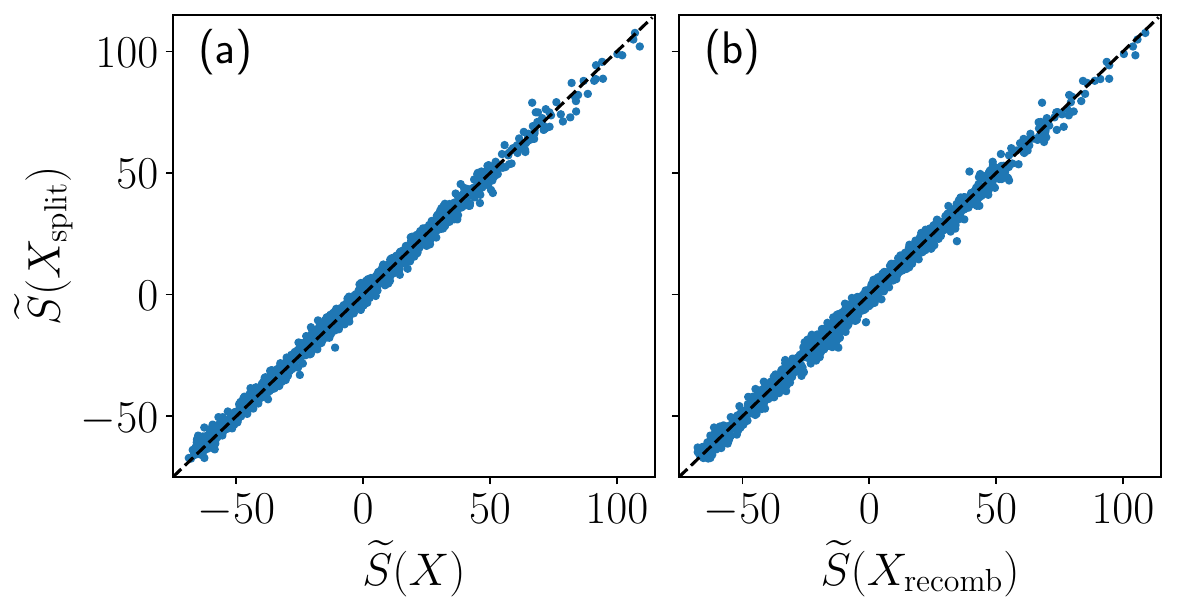}
  \caption{
    Demonstration of \ref{A5:splitting}.
    The representation $\tilde{S}$ of a split state $X_{\mathrm{split}}$ is plotted against $\tilde{S}(X)$ in (a) and against $\tilde{S}\bigl(X_{\mathrm{recomb}}\bigr)$ in (b).
    The dashed lines show $\tilde{S}\bigl(X_{\mathrm{split}}\bigr) = \tilde{S}(X)$ in (a) and $\tilde{S}\bigl(X_{\mathrm{split}}\bigr) = \tilde{S}\bigl(X_{\mathrm{recomb}}\bigr)$ in (b), respectively.
  }
  \label{fig:axiom5}
\end{figure}

Fifth, we consider the splitting and recombination \ref{A5:splitting}.
The method of creating a split state $X_{\mathrm{split}} = \bigl(\lambda X, (1 - \lambda) X\bigr)$ from a state $X$ is as follows. We insert a potential wall that is parallel to either the x-direction (at $L_y^{\mathrm{wall}}=\lambda L$) or the y-direction (at $L_x^{\mathrm{wall}} = \lambda L_x$) into the state $X$.
We remove the particles that are located in the region of the wall to avoid the extremely large force between the wall and the particles.
We then perform the MD simulation from this new state for a short time with the wall and piston fixed to obtain a split state.
Since the inserted wall is a potential wall, there is no thermal contact between the subsystems of the split state.
We also create a recombined state $X_{\mathrm{recomb}}$ from a split state $\bigl(\lambda X, (1 - \lambda) X\bigr)$ by removing the wall and performing the MD simulation for a short time.
We investigate the representation $\tilde{S}$ of the original state $X$ and the split state $X_{\mathrm{split}}$ and the recombined state $X_{\mathrm{recomb}}$.
As shown in Fig.~\ref{fig:axiom5}, $\tilde{S}\bigl(X_{\mathrm{split}}\bigr)$ is close to both $\tilde{S}(X)$ and $\tilde{S}\bigl(X_{\mathrm{recomb}}\bigr)$ for most of the test data.
The determination coefficients $R^2$ of the plots in Figs.~\ref{fig:axiom5}(a) and (b) are $0.9949$ and $0.9946$, respectively, which demonstrates \ref{A5:splitting}.

\begin{figure}[tb]
  \centering
  \includegraphics[width=.82\linewidth]{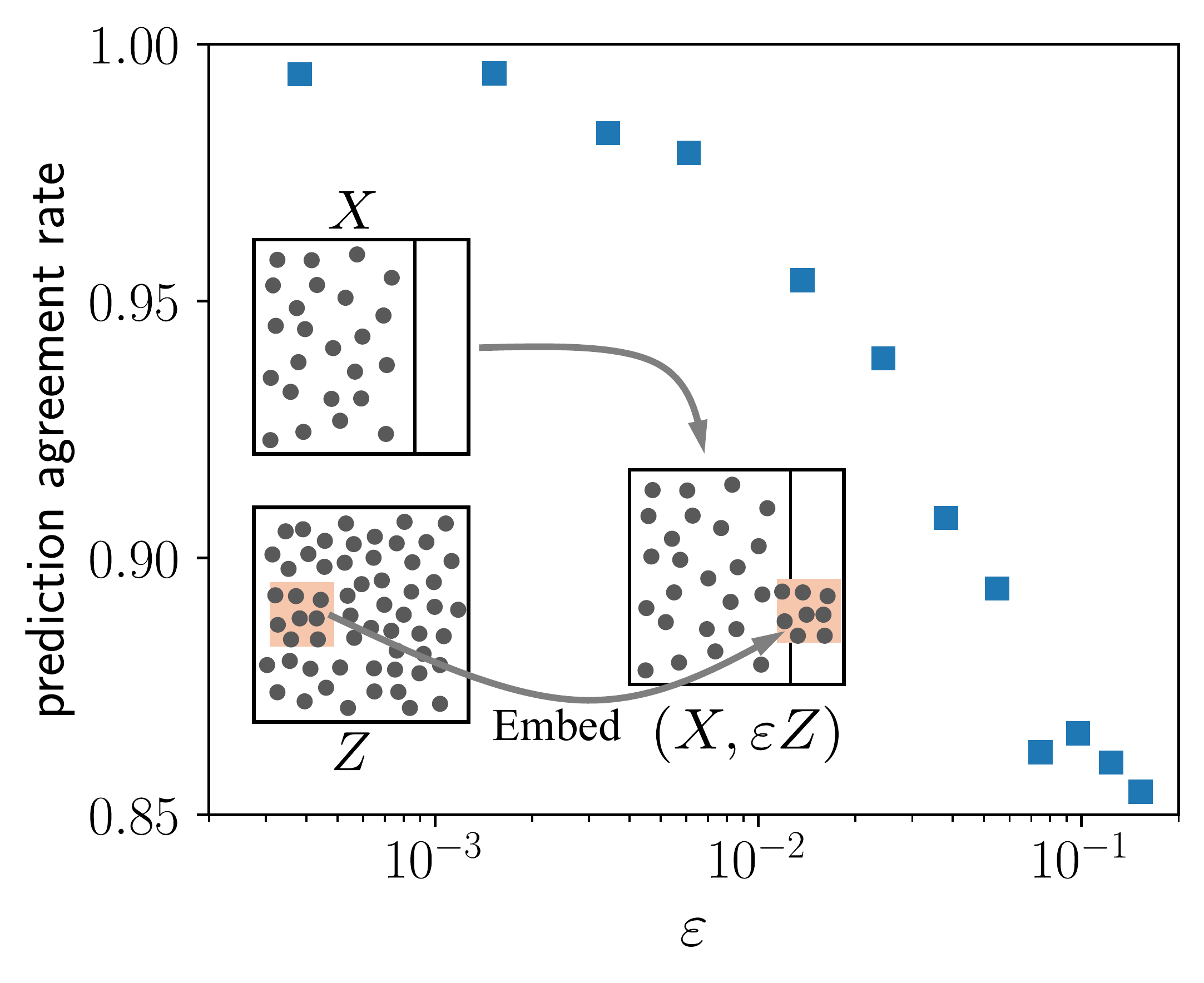}
  \caption{
    Demonstration of \ref{A6:stability}.
    The prediction agreement rate $| D^{\mathrm{agree}}_{\mathrm{A6}} | / | D_{\mathrm{A6}} |$ is plotted against the relative size $\varepsilon$ of a dust state $\varepsilon Z$.
    The inset shows how to create a state of $(X, \varepsilon Z)$ from $X$ and $Z$.
  }
  \label{fig:axiom6}
\end{figure}

Finally, we consider the stability \ref{A6:stability}.
With the help of the Comparison Hypothesis, when we restrict $X$ and $Y$ to belong to the same state space $\Gamma$,
the statement of \ref{A6:stability} is reduced to the following:
If $X \prec_{\mathrm{DNN}} Y$, then for sufficiently small $\varepsilon > 0$,
$(X, \varepsilon Z_1) \prec_{\mathrm{DNN}} (Y, \varepsilon Z_2)$ holds for any states $Z_1$ and $Z_2$.
That is, small dusts, $\varepsilon Z_1$ and $\varepsilon Z_2$, do not break the order relation between $X$ and $Y$.
Here we demonstrate this reduced statement.
We create a state $(X, \varepsilon Z)$ from $X$ and $Z$ by taking an $\varepsilon$-fraction of $Z$ and embedding it into $X$ (inset of Fig.~\ref{fig:axiom6}).
We investigate the prediction agreement rate, $| D^{\mathrm{agree}}_{\mathrm{A6}} | / | D_{\mathrm{A6}} |$,
where $D_{\mathrm{A6}} = \{(X, Y) \mid X \prec_{\mathrm{DNN}} Y \}$ and
$D_{\mathrm{A6}}^{\mathrm{agree}} = \{(X, Y) \in D_{\mathrm{A6}} \mid (X, \varepsilon Z_1) \prec_{\mathrm{DNN}} (Y, \varepsilon Z_2) \}$ with a fixed pair of states, $Z_1$ and $Z_2$.
Figure~\ref{fig:axiom6} shows that the prediction agreement rate approaches 1 as $\varepsilon$ decreases for a pair of $Z_1$ and $Z_2$.
We observe similar behaviors of the prediction agreement rate for several other pairs (not shown), which demonstrates \ref{A6:stability}.

\subsection{Representation $\tilde{S}$ as an entropy}
\label{subsec:representation_entropy}

In the previous subsection, we have shown that
the DNN-induced order relation $\prec_{\mathrm{DNN}}$ satisfies the axioms \ref{A1:reflectivity}--\ref{A6:stability} within the numerical precision.
We next demonstrate that the representation $\tilde{S}(X)$ of the trained DNN machine can be interpreted as a canonical entropy.
This demonstration justifies that $\prec_{\mathrm{DNN}}$ is actually identified with the order relation $\prec$ of the adiabatic accessibility
because $\prec_{\mathrm{DNN}}$ is determined based on $\tilde{S}(X_0)$ and  $\tilde{S}(X_1)$.

To demonstrate this, we numerically compare $\tilde{S}$ for a simple (non-compound) system of each fixed $N$ with the entropy of the van der Waals gas.
As described in Sec.~\ref{sec:md}, the system in the MD simulation is a two-dimensional gas of particles that interact only repulsively with each other.
Therefore, it is expected that its thermodynamic property is well described by the two-dimensional monatomic van der Waals gas without attractive interaction.
The thermodynamic entropy $S_{\mathrm{vdW}}(U, V, N)$ of the van der Waals gas of energy $U$, volume $V$, and number of particles $N$ reads \cite{Callen1985}:
\begin{align}
  S_{\mathrm{vdW}}(U, V, N)
   & = N k_{\mathrm{B}} \log \left[ \Bigl( u + \frac{a}{v} \Bigr)^c ( v - b ) \right] + N s_0
  \notag
  \\
   & = N k_{\mathrm{B}} \log \bigl[ u ( v - b ) \bigr] + N s_0,
  \label{vdW_entropy}
\end{align}
where $u = U/N$, $v = V/N$, and $s_0$ is a constant.
In the second line, we have used $a=0$ (because of no attractive interaction)
and $c=1$ (because of two-dimensional monatomic gas).
The constant $b$ represents the excluded volume per particle due to the repulsive interaction.
Since the sufficiently large interaction strength $C_{\mathrm{int}}$ is used in the MD simulation, it is reasonable to set $b = 2 \pi R^2$.

\begin{figure}[tb]
  \centering
  \includegraphics[width=.95\linewidth]{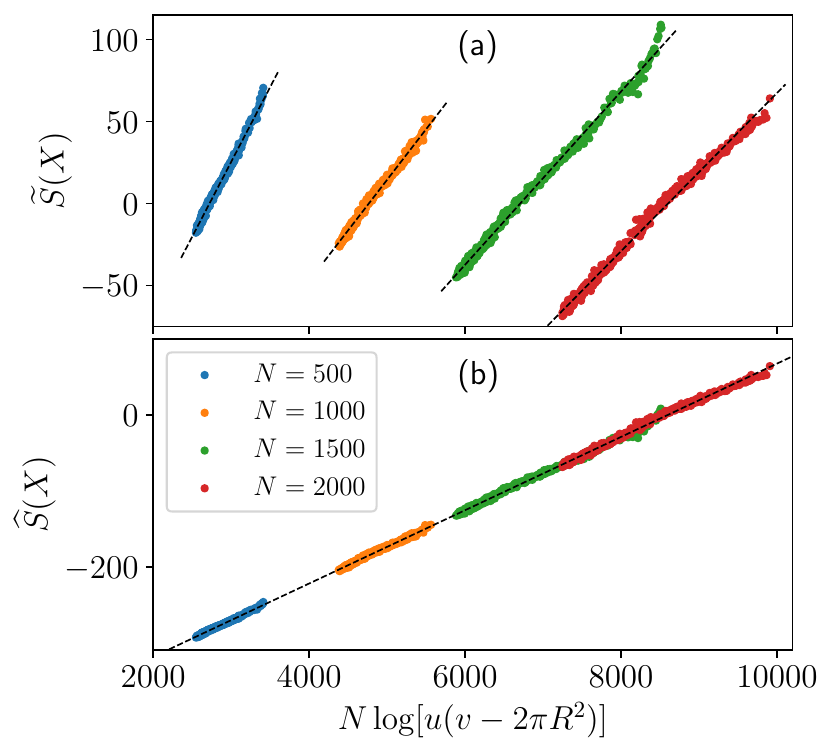}
  \caption{
    Parametric plots of (a) the original representation $\tilde{S}(X)$
    and (b) an appropriately affine transformed representation $\hat{S}(X)$
    against $N \log \bigl[ u(X) \bigl\{ v(X) - 2 \pi R^2 \bigr\} \bigr]$ for $N = 500, 1000, 1500, 2000$.
    The dashed lines are the fitted lines that are linearly dependent on $N \log \bigl[ u \bigl\{ v - 2 \pi R^2 \bigr\} \bigr]$.
  }
  \label{fig:vdW}
\end{figure}

In Fig.~\ref{fig:vdW}(a), we plot $\tilde{S}(X)$ against $N \log \bigl[ u(X) \bigl\{ v(X) - 2 \pi R^2 \bigr\} \bigr]$ for each of $N = 500, 1000, 1500, 2000$.
In this plot, each data point corresponds to a microscopic state $X$ obtained from the MD simulations.
For a state $X$, we compute the energy $U(X)$ [using Eq.~\eqref{energy_md}] and the volume $V(X) = LL_x(X)$ [$L_x(X)$ is the piston position at which $X$ is realized in the MD simulation].
We then compute $N \log [u(X) (v(X) - 2 \pi R^2)]$ with $u(X) = U(X)/N$ and $v(X) = V(X)/N$, which is an x-axis value of the plot.
The corresponding y-axis value is the neural network output $\tilde{S}(X)$ for the same state $X$.

We observe that $\tilde{S}(X)$ is linearly dependent on $N \log \bigl[ u(X) \bigl\{ v(X) - 2 \pi R^2 \bigr\} \bigr]$ for each $N$.
This implies that for each $N$,
$\tilde{S}$ is a function of only the two macroscopic variables, $u$ and $v$,
and it is an affine transformation of $S_{\mathrm{vdW}}$.
Therefore, we can interpret $\tilde{S}(X)$ as a canonical entropy $S_{\Gamma_N}(X | X_{*}^{(N)}, X_{**}^{(N)})$ for each $N$,
where $\Gamma_N$ is the state space for the system of $N$ particles and $X_{*}^{(N)}$ and $X_{**}^{(N)}$ are a pair of reference states in $\Gamma_N$.

Therefore, we can construct a thermodynamic entropy that is consistent across the different $N$ by an appropriate affine transformation $\hat{S}(X) = \alpha_N \tilde{S}(X) + \beta_N$.
We show an example of the transformation in Fig.~\ref{fig:vdW}(b),
where $\hat{S}(X) \propto S_{\mathrm{vdW}}(U, V, N)$ [Eq.\eqref{vdW_entropy}] with $s_0=0$.

We can also estimate the reference states $X_*^{(N)}$ and $X_{**}^{(N)}$ as follows.
From Eq.~\eqref{canonical_entropy}, $X_*$ and $X_{**}$ satisfy $S_\Gamma(X_* | X_*, X_{**}) = 0$ and $S_\Gamma(X_{**} | X_*, X_{**}) = 1$, respectively.
From Fig.~\ref{fig:vdW}(a), we have found the functional form of $\tilde{S}$ for fixed $N$ as $\tilde{S}_N(U,V) = a_N N \log [u (v - 2 \pi R^2)] + b_N$, where $a_N$ and $b_N$ are the fitting parameters.
We thus obtain the relation between the energy $U_*^{(N)}$ and volume $V_*^{(N)}$ of $X_*^{(N)}$ as $\tilde{S}_N(U_*^{(N)}, V_*^{(N)}) = 0$.
If we fix the value of either $U_*^{(N)}$ or $V_*^{(N)}$, we can determine the other:
for example, if we choose $V_*^{(N)} = L^2$, we obtain $U_*^{(N=500)}=666.1$, $U_*^{(N=1000)}=1402.3$, $U_*^{(N=1500)}=2443.5$, and $U_*^{(N=2000)}=3813.2$.
Once we determine $U_*^{(N)}$ and $ V_*^{(N)}$, we can generate a corresponding microscopic state $X_*^{(N)}$ from a MD simulation under given $U_*^{(N)}, V_*^{(N)}, N$.
In a similar manner, we also obtain $X_{**}^{(N)}$ and the corresponding energy $U_{**}^{(N)}$ and volume $V_{**}^{(N)}$.

Furthermore, we investigate the monotonicity, extensivity and additivity of $\tilde{S}$.
The monotonicity is valid
if the order relation induced by the trained DNN satisfies \ref{A1:reflectivity}--\ref{A6:stability}
because the DNN model is written as
$F_{\mathrm{DNN}}(X_0, X_1) = \Theta\bigl(\tilde{S}(X_0) - \tilde{S}(X_1) \bigr)$
with the step function $\Theta(x)$.
\begin{figure}[tb]
  \centering
  \includegraphics[width=.75\linewidth]{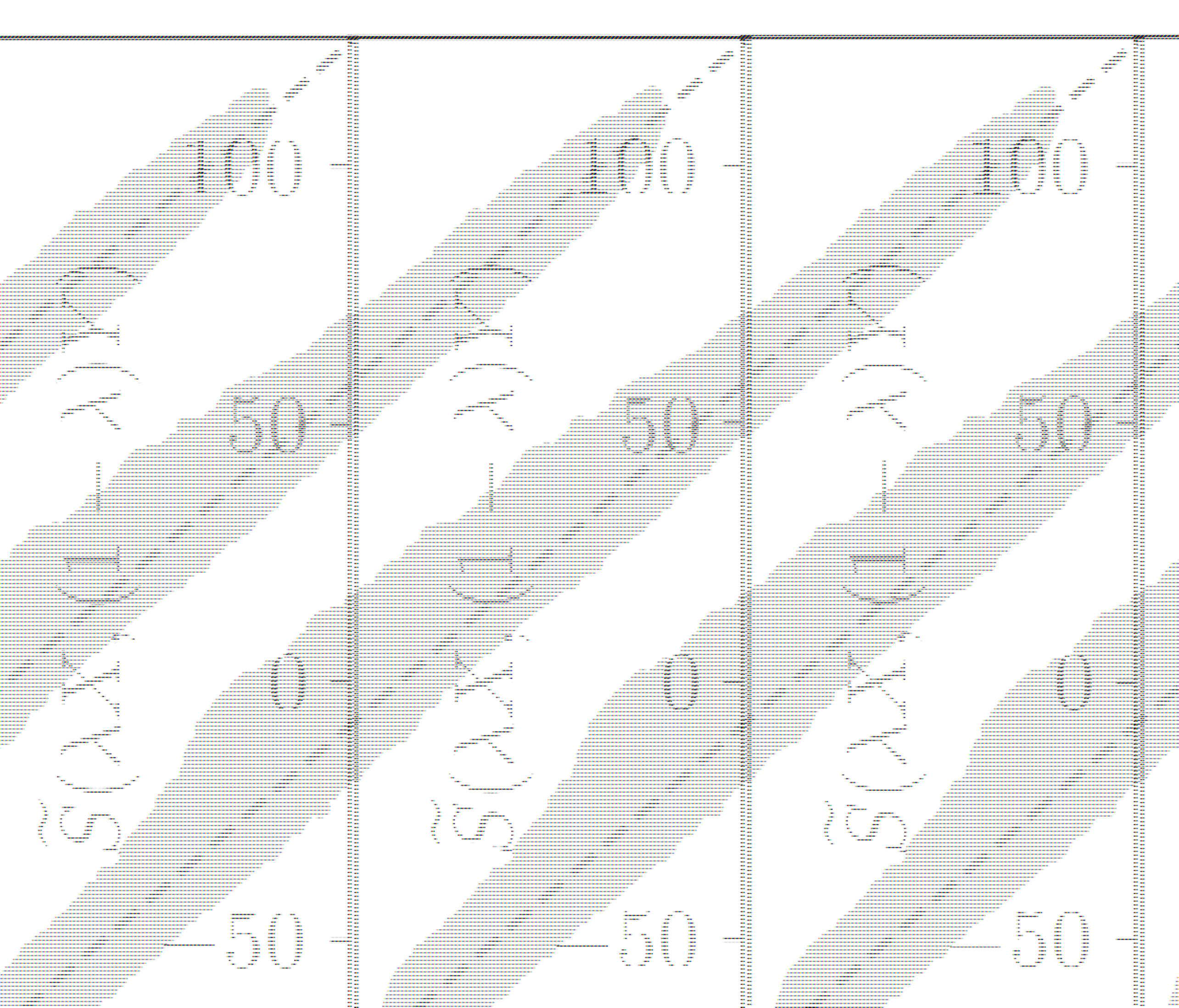}
  \caption{
    Representation $\tilde{S}$ of a compound state $\bigl(\lambda X, (1 - \lambda) Y \bigr)$
    plotted against the sum of the scaled representations, $\lambda  \tilde{S}(X) + (1 - \lambda)\tilde{S}(Y)$.
    The dashed line shows $\tilde{S}\bigl(\lambda X, (1 - \lambda) Y\bigr) = \lambda \tilde{S}(X) + (1 - \lambda) \tilde{S}(Y)$.
  }
  \label{fig:ext_add}
\end{figure}
We demonstrate the extensivity and additivity as follows.
We create a compound state $\bigl(\lambda X, (1 - \lambda) Y \bigr)$ of two scaled states $\lambda X$ and $(1 - \lambda) Y$ in a similar way to that in Fig.~\ref{fig:compound_scaling} with $\lambda=0.2, 0.3, ..., 0.8$.
In Fig.~\ref{fig:ext_add}, we compare $\tilde{S}\bigl(\lambda X, (1 - \lambda) Y \bigr)$ with $\lambda  \tilde{S}(X) + (1 - \lambda)\tilde{S}(Y)$,
where $X$ and $Y$ have the same number of particles.
We observe that $\tilde{S}\bigl(\lambda X, (1 - \lambda) Y \bigr)$ is almost linearly dependent on $\lambda  \tilde{S}(X) + (1 - \lambda)\tilde{S}(Y)$.
The coefficient of determination $R^2$ of the plot is $0.9766$, which illustrates that $\tilde{S}$ is an extensive and additive quantity.
Compared with the results for \ref{A1:reflectivity} (Fig.~\ref{fig:axiom1}) and \ref{A5:splitting} (Fig.~\ref{fig:axiom5}),
the deviation $\tilde{S}\bigl(\lambda X, (1 - \lambda) Y\bigr) - \bigl[ \lambda \tilde{S}(X) + (1 - \lambda) \tilde{S}(Y) \bigr]$ is larger and the coefficient of determination $R^2$ is smaller.
This would be due to the same reason as that of the lower prediction agreement rate of \ref{A4:scaling}.

\section{Summary}
\label{sec:summary}

In this study, we demonstrated that a DNN machine can behave as a Lieb--Yngvason machine---which is equipped with the order relation satisfying the axioms \ref{A1:reflectivity}--\ref{A6:stability} of thermodynamics---through the training with microscopic data of gas particles and without prior knowledge of thermodynamics.
The order relation is determined by the representation $\tilde{S}$ of the microscopic image extracted by the trained DNN, and it is interpreted as a canonical entropy of the state.

The effectiveness of our approach can be understood as the result of three complementary factors: data preparation, task design, and network architecture.

In terms of data preparation, an important element was that before sampling, we halted the piston and allowed the system to relax to an equilibrium state.
This procedure ensured that the collected data implicitly incorporated macroscopic irreversibility through equilibration.
Another crucial aspect was that the training data were obtained at various volumes.
This prevented the network from merely capturing a trivial monotonic increase of energy and instead enabled the emergence of entropy as the relevant quantity.
For example, had we restricted ourselves to snapshots at a fixed volume (e.g., $V=L^2$), the trajectories would have shown only monotonic energy increase [see the dips in Fig.~\ref{fig:profiles}(b)], in which case the learned representation would likely have been energy rather than entropy.

Regarding task design, as discussed in Sec.~\ref{sec:nn}, we trained the network to discriminate temporal order rather than directly compare adiabatic accessibility.
Because the data arose from adiabatic processes, temporal order naturally encoded the information on adiabatic accessibility.
This again highlights the crucial role of irreversibility embedded in the data.

Finally, the network architecture was tailored to the thermodynamic context: the representation extracted by the subnetwork was restricted to a single scalar value $\tilde{S}$, reflecting the fact that adiabatic accessibility in thermodynamics is characterized by a single quantity, entropy.

Each of these design choices was informed---albeit implicitly---by thermodynamic expectations.
Our results should therefore be understood as obtained without \textit{explicit} prior knowledge of thermodynamics.
We also note that while these factors are believed to have contributed to the success of our approach, the underlying reasons remain nontrivial and are not yet fully understood.

In our demonstration, we used a gas system with short-range repulsive interactions.
To establish the generality of our results, further investigations on a wider class of systems are necessary.
Possible directions include a system of Lennard--Jones particles \cite{Watanabe_etal2012}, which incorporate attractive interactions, and anharmonic lattice models \cite{SaitoDhar2010}.
Extending the analysis to quantum systems, beyond the classical regime considered here, is also an important future challenge.

Moreover, our study focused on densities well below the threshold for the Alder-type phase transition.
It would be valuable to examine systems that exhibit phase transitions or phase coexistence.
When the system enters the solid phase beyond the transition point, particle motion becomes more constrained, which may hinder the extraction of velocity information from snapshots. As a result, learning in the solid phase could be more challenging than in the gaseous phase.
Moreover, entropy exhibits singular behavior at the transition point in thermodynamics.
Thus, if we train a DNN on data that include adiabatic processes crossing the transition point, it would acquire an entropy-like quantity that reflects such singular features.

We also note that the present DNN was not designed to compare states with different particle numbers.
This reflects a property inherited from the Lieb--Yngvason machine, which admits only comparable states (in the sense of the Lieb--Yngvason framework) as input.
Since states with different particle numbers are not comparable---adiabatic operations cannot change the particle number---the output of our DNN is undefined when such states are provided as input.
As a result, the network output $\tilde{S}$ appears as different functions of $U, V$ for each $N$, as shown in Fig. 8(a).
However, after an appropriate affine transformation, we obtain $\hat{S}$, which collapses onto a single line valid for all $N$.
A possible way to directly address this indeterminacy would be to extend the temporal-order discrimination task into a three-class classification problem by introducing an additional category, i.e.,  ``incomparable.''

Another important direction is to explore nonequilibrium settings.
Whether macroscopic theories such as nonequilibrium thermodynamics \cite{LiebYngvason2013} or hydrodynamics can emerge from microscopic data remains an intriguing question.

Our results showed that entropy emerges in the representation of the DNN corresponding to an input of a microscopic state.
This suggests that the DNN performs a certain form of statistical mechanics.
However, unlike the Boltzmann principle, it does not involve counting over all microscopic states since the input is a single microscopic state.
This raises the possibility of a novel statistical mechanical framework.
Exploring this direction may provide new insights into the foundations of statistical mechanics.

\end{document}